\documentclass[aps,showpacs,twocolumn,footinbib]{revtex4}%
\usepackage{amssymb}
\usepackage{amsmath}
\usepackage{amscd}
\usepackage{graphicx}
\usepackage{subfigure}
\usepackage{float}

\begin{document}

\title{Phase-covariant cloning via adiabatic passage in fiber-nanocavity system}
\author{Wan-Jun Su}
\email{wanjunsu@fzu.edu.cn}
\affiliation{Department of Physics, Fuzhou University, Fuzhou
350002, People's Republic of China}
\affiliation{ Institute for Quantum Information Science, University of Calgary, Alberta T2N 1N4, Canada}
\author{Zhen-Biao Yang}

\affiliation{Department of Physics, Fuzhou University, Fuzhou
350002, People's Republic of China}


\date{\today}

\begin{abstract}
We propose an effective scheme for realizing a long-range quantum state phase-covariant cloning between two qubits in fiber-nanocavity system via an adiabatic passage. Since no cavity (fiber) photons or excited levels of the nitrogen-vacancy (NV) center are populated during the whole process, the scheme is immune to the decay of cavity (fiber) and spontaneous emission of the NV center. The strictly numerical simulation shows that the fidelity is high even in the presence of realistic imperfections.
\end{abstract}

\pacs{03.65.Xp,03.65.Vf,42.50.Dv,42.50.Pq}
\keywords{phase-covariant cloning,adiabatic passage,nitrogen-vacancy center, nanocavity }

\maketitle

\section{Introduction }

It is well known that the fundamental for quantum information
science is no-cloning theorem, which states that an unknown quantum
state cannot be cloned perfectly \cite{W1982}. However, we can try to
achieve a state as close as possible to the input state. Recently,
different schemes of quantum cloning have been proposed, and various
quantum cloning machines were designed for different tasks
\cite{R1998,V1998}. For the universal quantum cloning machine (UQCM), the
input can be arbitrary qubits, the fidelity is optimal and does not
depend on the input qubit \cite{Buzek1996}. It has been reported in the
NMR system \cite{Cummings2002} and linear optics system \cite{Huang2001,A2002}.
Different form UQCM, phase-covariant quantum cloning machine \cite{D2000}
restricts the input state of the equatorial qubit
$|\psi\rangle=(|0\rangle+e^{i\phi}|1\rangle)/\sqrt{2}$, which is
located on the equator of the Bloch sphere. Phase-covariant quantum
cloning machine has been reported in the solid state system with a
single electron spin at room temperature \cite{Pan2011}, and the corresponding fidelity  even reached $85.4\%$.

The key of quantum cloning is entanglement. Especially, tripartite entanglement
states have been used to realize quantum cloning. The W state is one important tripartite entangled state, which was proposed by D$\ddot{u}$r and has some interesting properties\cite{Dur2000}. In recent experiments, it was shown that three-qubit W states can be realized in optical systems \cite{Eibl2004} and ion
traps \cite{Roos2004}. Since W state retains bipartite entanglement when any one of the three qubits is traced out, it can be used in quantum
information processing \cite{Dr1998} and to test quantum nonlocality
without inequality \cite{Zheng2002,Cabello2002}. What's more, the schemes using W states to realize phase-covariant quantum cloning have been studied in various physical systems \cite{Zheng2005,Shen2012}.

The main obstacle to realizing multi-particle entanglement and
quantum information processing is decoherence. In the process of
decoherence in cavity QED, atomic spontaneous emission and cavity
decay take effect. How to reduce the decoherence is an important
problem. Adiabatic techniques are the answers since they feature a
certain robustness, and in systems of $\bigwedge$ type one can avoid
a transient large population in the excited state. Recently, the
techniques of stimulated Raman adiabatic passage \cite{K1998}
and fractional stimulated Raman adiabatic passage \cite{N1999}
have been extensively used for realizing the quantum information
processing \cite{X2006,Yang2010,Song2010}.

Motivated by these works, we present a new scheme to generate W
state of NV centers in fiber-cavity coupling system via the adiabatic passage.
Meanwhile, we realize phase-covariant cloning using the W state. In this scheme, we use the ground states of NV centers, and the cavities (fibers) field remains in the vacuum
state throughout the operation. So NV center's spontaneous emission and cavity decay
can be efficiently suppressed. Another advantage of the scheme is
that the interaction time does not need to be accurately controlled as
long as the adiabatic condition is fulfilled, which makes it more
applicable to current laboratory techniques. Our scheme may offer
promising perspectives for entanglement generation and quantum
information processing.

\section{Physical model}

\begin{figure}[tbp]
 \centering
  \includegraphics[width=0.4\textwidth]{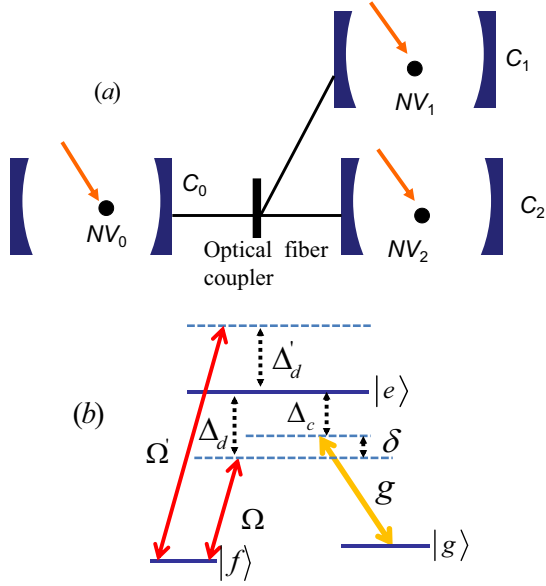} \caption{(Color online)
(a)The experimental setup for generating W states of NV centers in fiber-nanocavity
coupled system (b) Configurations of the NV center level structure
and relevant transitions. }
\end{figure}
The experimental setup for generating W state of NV centers in the
coupled system, as shown in Fig. 1(a). Three NV centers (labeled
$NV_0$, $NV_1$, $NV_2$) are separately trapped in three distant
optical cavities (labeled $C_0$, $C_1$, $C_2$) via an optical fiber
coupler. The fiber coupler connects three fibers, at the end of
which connect three cavities. The ground state of NV center is a
spin triplet, which labeled as $^3A$. There is a zero-field
splitting (2.88GHz) with a zero-field splitting between the state
$\left| 0\right\rangle $ ($m_s=0$) and $\left| \pm 1\right\rangle
$ ($m_s=\pm1$) \cite{Togan2010}. When an external
magnetic field are added along the NV center symmetry axis, the levels
$\left| \pm 1\right\rangle $ can be split by $2\pi\times200$ MHz,
while the $\left| 0\right\rangle $ state was not affected.  For
simplicity, we denote the following states:
$\left|0\right\rangle=\left| g\right\rangle$ and
$\left|-1\right\rangle=\left|f\right\rangle $ are the ground states.
while $\left|1\right\rangle=\left|e\right\rangle$ is the excited
state. As shown in Fig. 1(b), the transition $\left| f\right\rangle
\leftrightarrow \left| e\right\rangle $ is derived by two classical
fields with the Rabi frequencies $\Omega_k$ and  $\Omega_k^{'}$,
and the corresponding detunings $\Delta_d$ and $\Delta_{d}^{'}$.
The transition $\left| g\right\rangle \leftrightarrow \left|
e\right\rangle $ is coupled to the cavity mode with the coupling
strength $g$, and the corresponding detuning is $\Delta_c$. In the
interaction picture, the Hamiltonian describing NV centers and
cavities interaction is ($\hbar=1$)
\begin{eqnarray}\label{1}
H&=&H_{NVc}+H_{fc},\cr
H_{NVc}&=&\sum_{k=0}^{2}[(\Omega_{k} e^{i\Delta_d
t}|e\rangle_{k}\langle f|+\Omega_{k}^{'} e^{-i\Delta_{d}^{'}
t}|e\rangle_{k}\langle f|\cr&&+ga_{k}e^{i\Delta_c
t}|e\rangle_{k}\langle g|)+H.c.],
\end{eqnarray}
where $a$ is the annihilation operator for the cavity mode. Under
the large detuning condition, i.e., $|\Delta_c|,|\Delta_d|,|\Delta_{d}^{'}|\gg
g,\Omega_{k},\Omega_{k}^{'}$, the upper level $|e\rangle$ can be adiabatically
eliminated. We set the parameters
$\Omega_{k}=\Omega_{k}^{'}$ and $\Delta_{d}=-\Delta_{d}^{'}$ to
eliminate the Stark shift induced by the classical lasers. The
cavities are initially in vacuum states, so the Stark shift induced
by the cavity mode is discarded. Furthermore, choosing the detunings
appropriately $\Delta_c=\Delta_d=\Delta$, the dominant Raman
transition is induced by the classical field and the cavity mode,
while the other Raman transitions are far off-resonant, and can be
neglected. Then, the Hamiltonian describing NV centers and cavities
interaction is rewritten as
\begin{eqnarray}\label{2}
H_{NVc}^{'}&=&\sum_{k=0}^{2}(\lambda_{k}a_{k}|f\rangle_{k}\langle
g|+H.c.),
\end{eqnarray}
where $\lambda_{k}=g\Omega_{k}/\Delta$ is the effective coupling
strength of the Raman transition induced by the classical field and
the cavity mode.

In this model, the optical cavities are connected by identical
single-mode fibers. We assume that all the fibers connected to the
coupler have the same transverse mode. In the short fiber limit,
$l\bar{\nu}/(2 \pi c)\leq1$ \cite{Serafini2006}, where $l$ is the length
of the fiber and $\bar{\nu}$ is the decay rate of the cavity fields
into the fiber modes, only one resonant mode $b$ of the
fiber interacts with the cavity modes. In the case, the coupling
between the fiber mode and cavity fields is modeled by the
interaction Hamiltonian
\begin{eqnarray}\label{3}
H_{fc}=\nu\sum_{k=0}^{2}( b^{+}a_{k}+H.c.),
\end{eqnarray}
where $\nu$ is coupling strength of the cavity mode to the fiber mode, and $b^{+}$ is the creation operator for the fiber mode. In the interaction picture, the total Hamiltonian describing NV centers, cavities and fibers interaction is given by
\begin{eqnarray}\label{4}
H_{total}=H_{NVc}^{'}+H_{fc},
\end{eqnarray}

\section{W state preparation and phase covariant cloning}

Now, we show how to generate W state of NV centers using the physical model above. For an initial state of the system $|fgg000\rangle|0\rangle_{f}$, the single excitation subspace
$\forall$ can be spanned by the following state vectors: $\{|\phi_0\rangle,|\phi_1\rangle,|\phi_2\rangle,|\phi_3\rangle, |\phi_4\rangle,|\phi_5\rangle,|\phi_6\rangle \}$, with
\begin{eqnarray}\label{5}
|\phi_0\rangle=|fgg000\rangle|0\rangle_{f},\cr
|\phi_1\rangle=|ggg100\rangle|0\rangle_{f},\cr
|\phi_2\rangle=|ggg000\rangle|1\rangle_{f},\cr
|\phi_3\rangle=|ggg010\rangle|0\rangle_{f},\cr
|\phi_4\rangle=|ggg001\rangle|0\rangle_{f},\cr
|\phi_5\rangle=|gfg000\rangle|0\rangle_{f},\cr
|\phi_6\rangle=|ggf000\rangle|0\rangle_{f}.
\end{eqnarray}
The Hamiltonian $H_{total}$ has the following dark state with null
eigenvalue as well
\begin{eqnarray}\label{6}
|\psi\rangle=\frac{1}{\sqrt{K}}(\frac{1}{\lambda_0}|\phi_0\rangle
+\frac{1}{\lambda_1}|\phi_5\rangle+\frac{1}{\lambda_2}|\phi_6\rangle-
\frac{1}{\nu}|\phi_2\rangle).
\end{eqnarray}
where $K=(1/{\lambda_{0}^{2}}+1/{\lambda_{1}^{2}}
+1/{\lambda_{2}^{2}}+1/{\nu^{2}})$ is the normalized factor.

According to the adiabatic theory, if the whole system is initially prepared in the dark state, under the adiabatic condition, it will evolve in the dark state subspace.
We control the Rabi frequencies of classical fields $\Omega_1=\Omega_2=\Omega$ to satisfy the effective coupling strengths $\lambda_1=\lambda_2$. At the beginning, the Rabi frequencies are assumed $\Omega\gg \Omega_{0}$. Then we slowly decrease $\Omega(t)$
and simultaneously increase $\Omega_0(t)$ to obtain $\Omega(t)/\Omega_{0}(t)=1$ at the time $T$. Meanwhile, assuming $\lambda_1=\lambda_0 \ll \nu$, we can discard the last term in the
Eq. (6) above. From the discussion above, the pulse shapes must satisfy
\begin{eqnarray}\label{7}
\lim\limits_{t \rightarrow -\infty}\frac{\Omega_{0}(t)}{\Omega(t)}=0,
\end{eqnarray}
and
\begin{eqnarray}\label{8}
\lim\limits_{t \rightarrow \infty}\frac{\Omega(t)}{\Omega_{0}(t)}=1.
\end{eqnarray}
As a result, we can obtain the target state
\begin{eqnarray}\label{9}
|\psi\rangle_{W}=\frac{1}{\sqrt{3}}(|fgg\rangle+|gfg\rangle+|ggf\rangle)|000\rangle_{c}|0\rangle_{f}.
\end{eqnarray}
 The Eq. (9) above denotes that the fiber and cavity field are in vacuum states, and the three NV centers are in three-particle W state.

\begin{figure}[tbp]
 \centering
  \includegraphics[width=0.4\textwidth]{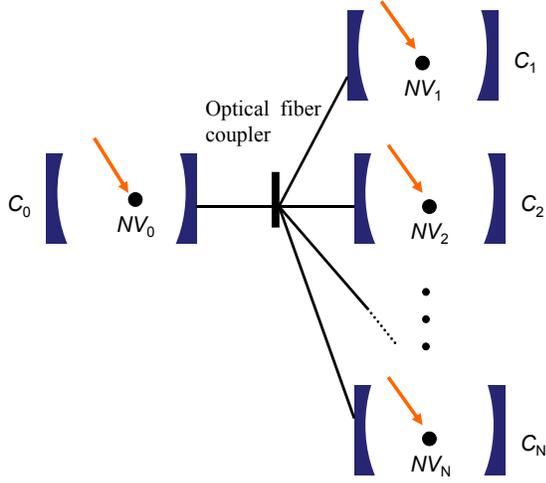} \caption{(Color online) Schematic of
$N$ ($N>3$) NV centers separately trapped in each of the distant cavities, which are connected by optical fibers. }
\end{figure}

Then, we show that the idea can be generalized to the generation of the multi-NV-centers W state. $N$ ($N>3$) NV centers are separately trapped in each of the distant cavities connected by fibers, as shown in Fig. 2. We assume that the system is initially in the state $|f_0g_1g_2\cdot\cdot\cdot
g_N\rangle|0\rangle_{c}\rangle|0\rangle_{f}$, where
$|0\rangle_{c}\rangle|0\rangle_{f}=\Pi_{k=0}^{N}|0_k\rangle_{c}|0_k\rangle_{f}$
denotes all the cavity fields and fiber fields are in the vacuum states.
Similar to the Eq. (6), we can get the dark state
\begin{eqnarray}\label{10}
|\psi\rangle&=&\frac{1}{\sqrt{K}}(\frac{1}{\lambda_0}|f_0g_1g_2\cdot\cdot\cdot
g_N\rangle|0\rangle_{c}\rangle|0\rangle_{f}+\cr&&\frac{1}{\lambda_1}|g_0f_1g_2\cdot\cdot\cdot
g_N\rangle|0\rangle_{c}\rangle|0\rangle_{f}+\cdot\cdot\cdot\cr&&+\frac{1}{\lambda_N}|g_0g_1\cdot\cdot\cdot
g_{N-1}f_N\rangle|0\rangle_{c}\rangle|0\rangle_{f}\cr&&
-\frac{1}{\nu}|g_0g_1g_2\cdot\cdot\cdot
g_N\rangle|0\rangle_{c}\rangle|1\rangle_{f}),
\end{eqnarray}
where $K=(1/{\lambda_{0}^{2}}+1/{\lambda_{1}^{2}} +\cdot\cdot\cdot+1/{\lambda_{N}^{2}}+1/{\nu^{2}})$ is
the normalized factor. Choosing $\Omega_1=\Omega_2=\cdot\cdot\cdot
\Omega_N=\Omega$, we get $\lambda_1=\lambda_2=\cdot\cdot\cdot
\lambda_N=\lambda$. At the beginning, we assume $\Omega\gg
\Omega_{0}$. Then we slowly decrease $\Omega(t)$ and simultaneously
increase $\Omega_0(t)$ to obtain $\Omega(t)/ \Omega_{0}(t)=1$ at the
time $T$.  Meanwhile, assuming $\lambda_0=\lambda \ll \nu$, we can
discard the last term in the Eq. (10) above. The final state is
\begin{eqnarray}\label{11}
|\psi\rangle_{W}&=&\frac{1}{\sqrt{N}}(|f_0g_1g_2\cdot\cdot\cdot
g_N\rangle+|g_0f_1g_2\cdot\cdot\cdot
g_N\rangle+\cdot\cdot\cdot\cr&&+|g_0g_1\cdot\cdot\cdot
g_{N-1}f_N\rangle).
\end{eqnarray}
So $N$ NV centers are prepared in an entangled state with the cavity
mode and fiber mode left in the vacuum states.

The quantum cloning scheme can be implemented based on the W state previously prepared. Again, we assume that all the cavities
and the optical fiber channel are initially both in the vacuum
state, and only one NV center of the first cavity is prepared in the
arbitrary orbital state of the Bloch sphere. The system initial
state is written as
\begin{eqnarray}\label{12}
|\psi_{(0)}\rangle&=&\frac{1}{\sqrt{2}}(|g_0\rangle+e^{i\delta}|f_0\rangle)|g_1g_2\cdot\cdot\cdot
g_N\rangle|0\rangle_{c}\rangle|0\rangle_{f}.
\end{eqnarray}

Under the above-mentioned conditions, the dark state
$|g_0g_1g_2\cdot\cdot\cdot
g_N\rangle|0\rangle_{c}\rangle|0\rangle_{f}$ undergoes no changes.
On the other hand, $|f_0g_1g_2\cdot\cdot\cdot
g_N\rangle|0\rangle_{c}\rangle|0\rangle_{f}$ evolves to the state of the Eq. (10).
So the system evolves to the state
\begin{eqnarray}\label{13}
|\psi_{(t)}\rangle&=&\frac{1}{\sqrt{2}}[|g_0g_1g_2\cdot\cdot\cdot
g_N\rangle+\frac{1}{\sqrt{N}}e^{i\delta}(|f_0g_1g_2\cdot\cdot\cdot
g_N\rangle\cr&&+|g_0f_1g_2\cdot\cdot\cdot
g_N\rangle+\cdot\cdot\cdot\cr&&+|g_0g_1\cdot\cdot\cdot
g_{N-1}f_N\rangle)|0\rangle_{c}\rangle|0\rangle_{f}].
\end{eqnarray}
All the cavity and fiber field are always in vacuum states, so this
scheme is robust against the cavity and fiber decay.

\section{discussion }

It should be noted that preparing W state of the NV centers is the
key step to achieving the phase-covariant quantum cloning. For this
reason, it is necessary to consider the feasibility of generating
the W state of the NV centers. In the above derivations, we
reasonably select the experimental parameters to satisfy the
adiabatic passage. To make the scheme experimentally feasible, we
discuss the Rabi frequencies and other correlative parameters. The
time-dependent pulses we introduce are as follows:
\begin{eqnarray}\label{14}
\Omega(t)&=&\Omega_{m}e^{-(t-t_{1})^2/t_p^{2}}+\Omega_{m}e^{-(t-t_{0})^2/t_p^{2}},\cr
\Omega_{0}(t)&=&\Omega_{m}e^{-(t-t_{0})^2/t_p^{2}},
\end{eqnarray}
where $\Omega_{m}$, $t_{k} (k=0,1)$ and $t_{p}$, are the peak, the time
delay and the width of the laser pulses, respectively. We utilize the fidelity and
the fidelity of the prepared states is given by
\begin{eqnarray}\label{15}
F=\left\langle \Psi _{(t)}|| \Psi _{ideal}\right\rangle,
\end{eqnarray}
where $\left| \Psi _{ideal}\right\rangle$ is the ideal three-NV centers W state in Eq. (10), and $\left| \Psi _{(t)}\right\rangle$ is the final state described
by the Schr\"{o}dinger equation $i (d\left| \Psi
_{(t)}\right\rangle/dt) =H\left| \Psi _{(t)}\right\rangle $, where
$H$ is the full Hamiltonian governed by Eq. (1).

\begin{figure}[tbp]
\centering
\subfigure
{\includegraphics[width=0.4\textwidth]{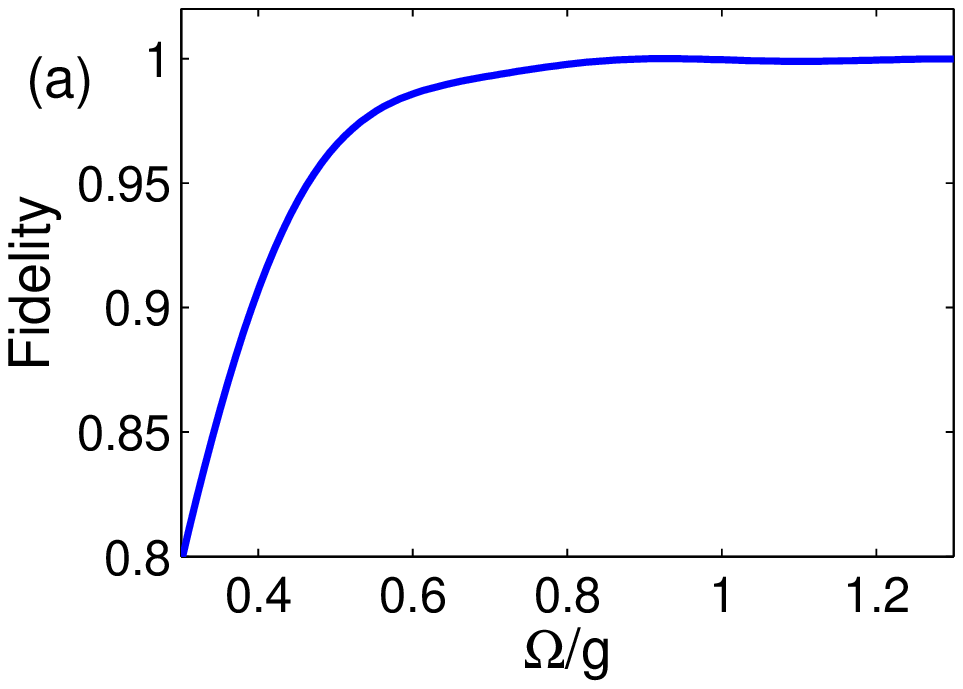}}
\subfigure
{\includegraphics[width=0.4\textwidth]{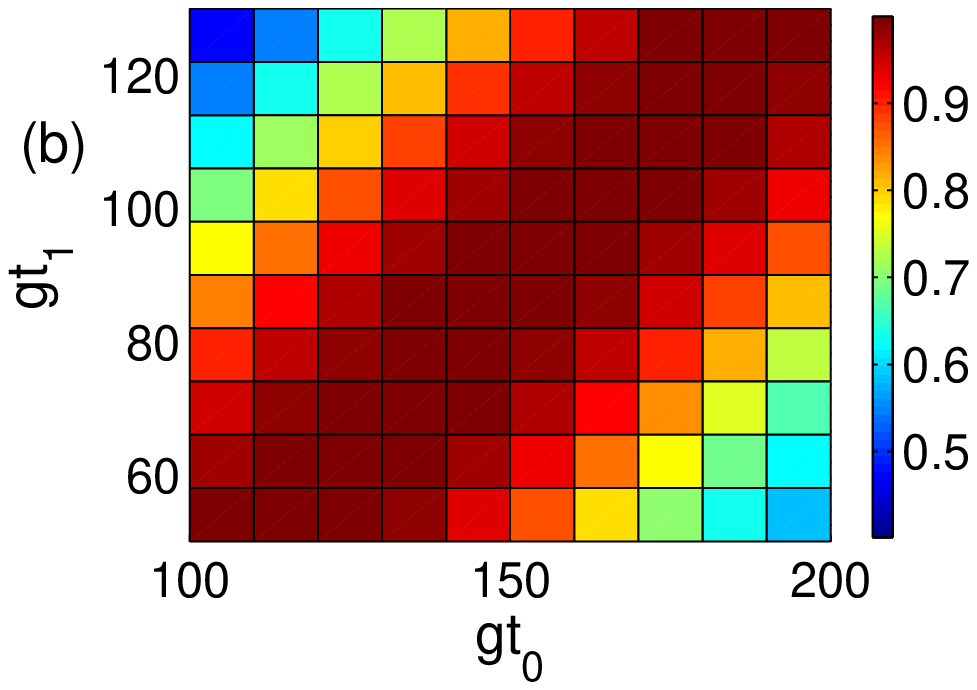}}
\subfigure
{\includegraphics[width=0.4\textwidth]{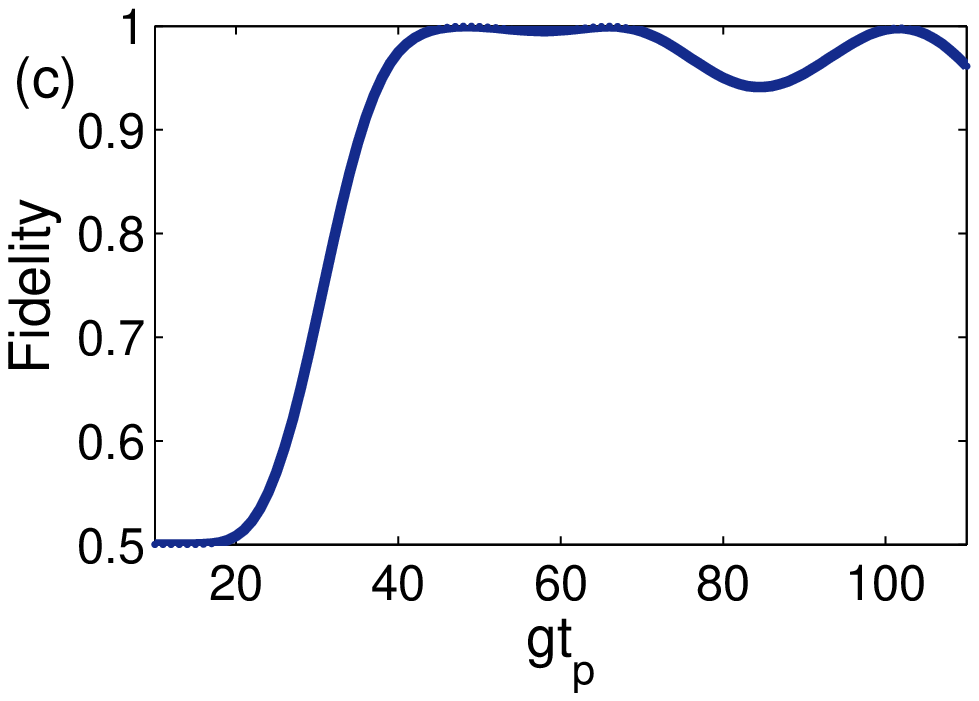}}
\caption{(Color online)Fidelity of  three-NV centers W state versus (a) the scaled peak of the laser pulses $\Omega_{m}/g$ with $gt_{p}=50$, $gt_{0}=150$ and $gt_{1}=90$; (b) the scaled time delay of the laser pulses $gt_{k}(k=0,1)$ with $\Omega_{m}=g$, $gt_{p}=50$; (c) the scaled width of the laser pulses $g{t_{p}}$ with $gt_{0}=150$ and $t_{1}=90g$ and $\Omega_{m}=g$. The other factors are assumed to be $\Delta=10g$, $\nu=10g$, $T=200/g$ in all the cases. }
\end{figure}

Fig. 3, we plot the dependences of the fidelity of three-NV centers W state versus the scaled parameters of the laser pulses, i.e, the scaled peak $\Omega_{m}/g$, the scaled time delay $gt_{k} (k=0, 1)$ and the scaled width $g{t_{p}}$. As we can see from the subfigures, when the parameters are set in a significant range, the fidelity reaches to unit one. The Fig. 3(a) shows that small $\Omega_{m}$ will reduce the fidelity seriously. Increasing the Rabi frequency $\Omega_{m}$ will get a raise in the effective coupling strength of the Raman transition induced by the classical field and the
cavity mode, and the effective Hamiltonian in Eq. (2) takes effect. The excited state can be eliminated, so the decoherence is reduced. We can see from the Fig. 3(b) that the fidelity decreases rapidly when $t_{0}-t_{1}$ is too large or too small. Obviously, when $t_{1}$ is close to $t_{0}$ or deviate much from $t_{0}$, the condition $\Omega(t)/\Omega_{0}t=1$ can not be satisfied at the end of evolution. In Fig. 3(c), we can find when $40<gt_{p}<60$, the fidelity is larger than 0.98. It is because the large deviation can not ensure the Eq. (7) and Eq. (8) to take effect. Hence, a high-fidelity W state can be obtained as long as the experiment parameters are set in the correspond range: $\Omega_{m}=g$, $gt_{0}=150$, $gt_{1}=90$ and $gt_{p}=50$.

\begin{figure}[tbp]
 \centering
  \includegraphics[width=0.4\textwidth]{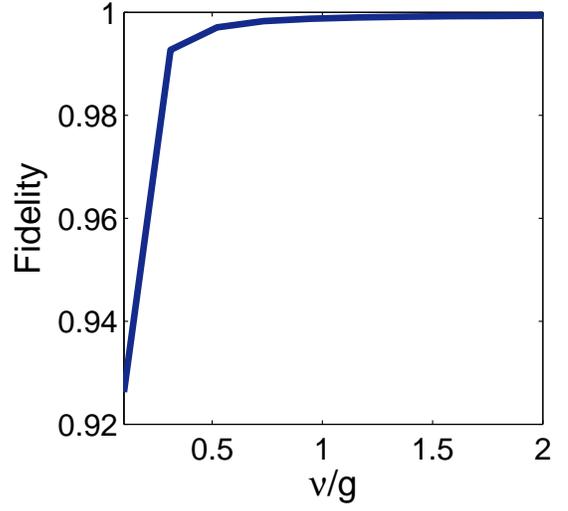} \caption{(Color online)
Fidelity of  three-NV centers W state versus the scaled cavity-fiber coupling strength $\nu/g$, with $\Delta=10g$, $\Omega_{m}=g$, $gt_{p}=50$, $gt_{0}=150$, $gt_{1}=90$,  $T=200/g$. }
\end{figure}

\begin{figure}[tbp]
 \centering
  \includegraphics[width=0.4\textwidth]{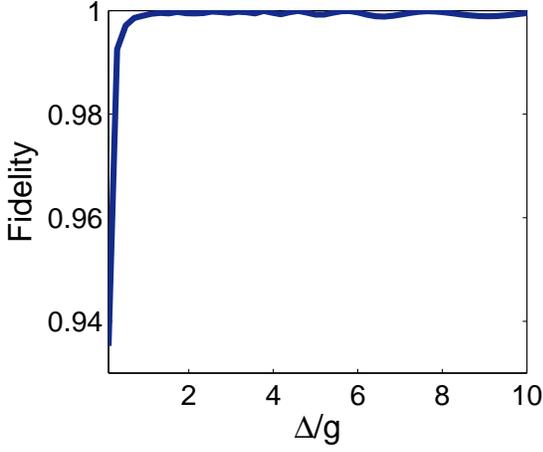} \caption{(Color online)
Fidelity of three-NV centers W state versus the scaled detuning $\Delta/g$, with $\nu=10g$, $\Omega_{m}=g$, $gt_{p}=50$, $gt_{0}=150$, $gt_{1}=90$, $T=200/g$. }
\end{figure}

\begin{figure}[tbp]
 \centering
 \includegraphics[width=0.4\textwidth]{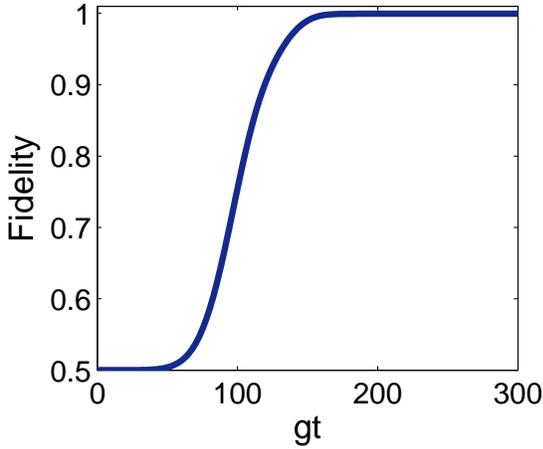} \caption{(Color online)
 Fidelity of three-NV centers W state versus the scaled interaction time $gt$, with $\Delta=10g$, $\nu=10g$, $\Omega_{m}=g$, $gt_{p}=50$, $gt_{0}=150$, $gt_{1}=90$.}
\end{figure}

To satisfy the condition of adiabatic passage $\nu\gg\Omega g /\Delta $, we should choose the coupling strength $\nu$ and the detuning $\Delta$ to be large enough. The Fig. 4 and Fig. 5 show that when $\nu/g>1$ and $\Delta/g>1 $, the fidelity of W state can reach 0.99. Considering the experimental feasibility, we choose $\nu=10g$ and $\Delta=10g$. Moreover, one observes from Fig. 6 that within pulse duration, the fidelity increases with the time going on and approximates unit 1 when $gt>150$. For the reason that the target state approaches a steady state at least, the interaction time need not to be controlled strictly, the same as the parameters. We choose $T/g=200$ as the evolution time with $\nu=10g$, $\Delta=10g$, $\Omega_{m}=g$, $gt_{0}=150$, $gt_{1}=90$ and $gt_{p}=50$. The numerical result shows a good agreement with the expected result.

\begin{figure}[tbp]
 \centering
  \includegraphics[width=0.4\textwidth]{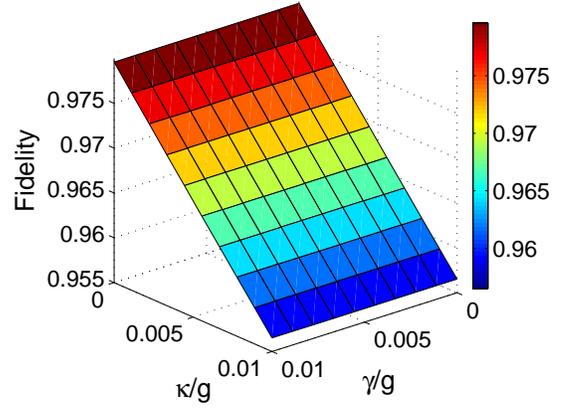} \caption{(Color online)
Fidelity of phase-covariant cloning using three-NV centers W state versus the scaled cavity(or fiber) decay rate $\kappa/g$  and atomic spontaneous emission rate $\gamma/g$, with $\Delta=10g$, $\nu=10g$, $\Omega_{m}=g$, $gt_{p}=50$, $gt_{0}=150$, $gt_{1}=90$, $T=200/g$. }
\end{figure}

Considering a realistic experiment,we must pay attention the effect of the spontaneous emission of the NV centers, fiber losses and cavity losses. The master equation of the whole system can be expressed by

\begin{eqnarray}\label{16}
\dot{\rho}&=&-i\left[H,\rho \right] +\frac{\kappa_c}2
(2a\rho a^{+}-aa^{+}\rho -\rho {a}^{+}a)\nonumber\\
&&+\frac{\kappa_f}2(2b\rho b^{+}-bb^{+}\rho -\rho {b}^{+}b)\cr&&+\sum_{j=g,f}\sum_{k=0}^{2}\frac{\gamma_k}2(2{\sigma
}_{je}\rho {\sigma }_{ej}-{\sigma }_{ej}{\sigma }_{je}\rho -\rho
{\sigma }_{ej}{\sigma }_{je}),
\end{eqnarray}
where $\kappa_c$ and $\kappa_f$ denote the effective decay rate of the cavity and the optical fiber. For simplicity, we assume $\gamma _{k}(k=0,1,2)=\gamma$, where
$\gamma$ represents the branching ration of the spontaneous decay
from level $\left| e\right\rangle $ to $\left| g\right\rangle $
and $\left| f\right\rangle $ in NV center. By solving the master equation
numerically, we obtain the relation of the fidelity of quantum phase-covariant cloning using three NV centers W states versus the scaled ratio $\gamma/g$ and
$\kappa/g$ with $\Delta=10g$, $\nu=10g$, $\Omega_{m}=g$, $gt_{p}=50$, $gt_{0}=150$, $gt_{1}=90$, $T=200/g$ in Fig. 7. We see that the fidelity is always larger than 0.955 even
$\gamma/g$ and $\kappa/g$ is up to 0.01.

We now want to study the performance of our protocol for quantum optical devices, such as microsphere cavity-fiber coupling system \cite{Park2006}. We can adjust the energy levels of different N-V centers using an external magnetic field to get identical N-V centers\cite{Togan2010}. When we put the NV centers near the equator of a microsphere cavity, where the NV center interacts with the cavity via the evanescent fields, we can get the coupling constants, which range from hundreds of MHz to several GHz \cite{Barclay2009}. In their work, they got a $Q$ factor of the microsphere cavity exceeded $2\times10^{6}$, which result in a photon leakage rate $\kappa=\omega/Q\sim 2\pi \times 120$ MHz \cite{Maze2008}. The spontaneous decay rate of the NV center has been reported is $\gamma\sim 2\pi\times 15$ MHz \cite{Santori2006}. Estimating a fiber-nanocavity system with the relevant cavity QED parameters $[g, \gamma, \kappa, \Omega, \Delta,\nu ]/2\pi= [1, 0.01, 0.01, 1, 10, 10]$ GHz, we can get the corresponding fidelity of the W state that can reach $95.5\%$. The operation time of the phase-covariant scheme is about $50ns$ with the parameters above. What's more, the decoherence time of individual NV centers is longer than $600\mu s$ at room temperature  \cite{Mizuochi2009}. During the decoherence time of NV center, we can complete the process of phase-covariant cloning. In principle, the performance of this scheme can be improved even further, by using shortcuts to adiabatic passage \cite{Chen2010}, which has been used in entangled state preparation and transition \cite{Ye-HongLPL2014}.

\section{conclusion}

In summary, based on adiabatic passage, we have
proposed a scheme for generating the $N$ NV-centers entangled state in a hybrid system consisting of NV centers, optical fibers, and micro-cavities. What's more, a scheme of phase-covariant cloning has been realized. By numerical calculation, we have demonstrated that the present scheme is immune to the excited levels and cavity (fiber) photons.

\section{acknowledge}
This work is supported by the National Fundamental Research Program People's Republic of China under Grant No. 11405031, and the Research Program of Fujian Education Department under Grant NO. JA14044, and the  Research Program of Fuzhou University.

\end{document}